\documentstyle[12pt,aaspp4,flushrt]{article}

\slugcomment{\it To appear in ApJS, Spitzer Special Issue}
\lefthead{Eisenhardt et al.}
\righthead{The IRAC Shallow Survey}

\def\eg{{e.g.,~}}
\def\etal{{et al.}}
\def\deg{\ifmmode {^{\circ}}\else {$^\circ$}\fi}

\def\ergcm2s{\ifmmode {\rm\,erg\,cm^{-2}\,s^{-1}}\else
     ${\rm\,erg\,cm^{-2}\,s^{-1}}$\fi}
\def\spose#1{\hbox to 0pt{#1\hss}}
\def\simlt{\mathrel{\spose{\lower 3pt\hbox{$\mathchar"218$}}
      \raise 2.0pt\hbox{$\mathchar"13C$}}}
\def\simgt{\mathrel{\spose{\lower 3pt\hbox{$\mathchar"218$}}
      \raise 2.0pt\hbox{$\mathchar"13E$}}}

\def\plotfiddle#1#2#3#4#5#6#7{\centering \leavevmode
\vbox to#2{\rule{0pt}{#2}}
\includegraphics{#1}}

\received{2004 April 8}
\begin{document}

\title
{The IRAC Shallow Survey}

\author
{P.~R.~Eisenhardt\altaffilmark{1},
D.~Stern\altaffilmark{1},
M.~Brodwin\altaffilmark{1},
G.~G.~Fazio\altaffilmark{2},
G.~H.~Rieke\altaffilmark{3},
M.~J.~Rieke\altaffilmark{3},
M.~W.~Werner\altaffilmark{1},
E.~L.~Wright\altaffilmark{4},
L.~E.~Allen\altaffilmark{2},
R.~G.~Arendt\altaffilmark{5},
M.~L.~N.~Ashby\altaffilmark{2},
P.~Barmby\altaffilmark{2},
W.~J.~Forrest\altaffilmark{6},
J.~L.~Hora\altaffilmark{2},
J.-S.~Huang\altaffilmark{2},
J.~Huchra\altaffilmark{2},
M.~A.~Pahre\altaffilmark{2},
J.~L.~Pipher\altaffilmark{6},
W.~T.~Reach\altaffilmark{7},
H.~A.~Smith\altaffilmark{2},
J.~R.~Stauffer\altaffilmark{7},
Z.~Wang\altaffilmark{2},
S.~P.~Willner\altaffilmark{2},
M.~J.~I.~Brown\altaffilmark{8},
A.~Dey\altaffilmark{8},
B.~T.~Jannuzi\altaffilmark{8},
and G.~P.~Tiede\altaffilmark{9}}

\altaffiltext{1}
{Jet Propulsion Laboratory,
California Institute of Technology,
MS 169-327, 4800 Oak Grove Drive, Pasadena, CA, 91109;
e-mail:  {\tt peisenhardt@sirtfweb.jpl.nasa.gov}}

\altaffiltext{2}{Harvard-Smithsonian Center for Astrophysics, 60 Garden
St., Cambridge, MA 02138}

\altaffiltext{3}{Steward Observatory, University of Ariziona, Tucson,
AZ, 85721}

\altaffiltext{4}{University of California, Los Angeles, CA 90095-1562}

\altaffiltext{5}{NASA Goddard Space Flight Center, Greenbelt, MD 20771}

\altaffiltext{6}{University of Rochester, Rochester, NY 14627}

\altaffiltext{7}{{\it Spitzer} Science Center, California Institute of
Technology, Pasadena, CA 91125}

\altaffiltext{8} {National Optical Astronomy Observatory, Tucson, AZ,
85726-6732}

\altaffiltext{9} {Bowling Green State University, Bowling Green, OH,
43403}

\begin{abstract}

The IRAC shallow survey covers $8.5\sq^\circ$ in the NOAO Deep
Wide-Field Survey in Bo\"otes with 3 or more 30 second exposures per
position.  An overview of the survey design, reduction, calibration, star-galaxy
separation, and initial results is provided.
The survey includes $\approx370,000,$ 280,000, 38,000, and 34,000 sources
brighter than the $5\sigma$ limits of 6.4, 8.8, 51, and 50 $\mu$Jy at
3.6, 4.5, 5.8, and 8 $\mu$m respectively,
including some with unusual spectral energy distributions.

\end{abstract}


\section{Introduction}

The million-fold lower background seen at infrared wavelengths in
space means that even brief exposures with a modest aperture telescope
like the {\it Spitzer Space Telescope} probe vastly larger volumes than
are possible from the ground.  For objects distributed uniformly in
Euclidean space, the number of sources detected is maximized by
observing a given field only long enough to become background limited,
and to reduce repositioning overheads to 1/3 of the observing time.
For uniform density sources observable to cosmological (non-Euclidean)
volumes, the advantage of short exposures is greater.  Such
considerations motivated the IRAC shallow survey ({\it Spitzer} program number 30), which detects sources
in the $L$-band $\approx 8,000 \times$ faster than did 
\markcite{Hogg:00}Hogg {et~al.} (2000)
using Keck, while reaching $5 \times$ deeper.  The 23 hour
{\it Infrared Space Observatory} observations at $6.7\mu$m of
\markcite{Sato:03}Sato {et~al.} (2003) detected sources a few times 
fainter than does the 90
second IRAC shallow survey at 5.8 and $8.0\mu$m, but at a rate several
hundred times slower.    The logical endpoint of this approach is an all-sky
survey such as that planned by the {\it Wide-field Infrared Survey
Explorer} \markcite{Eisenhardt:03}(Eisenhardt \& Wright 2003).

A major scientific driver for the IRAC shallow survey is the detection
of galaxy clusters at $z > 1$ via the redshifted $1.6\mu$m peak in
galaxy spectral energy distributions (SEDs).   Extending the evolution
observed in the $K$-band in clusters to $z\sim1$ 
\markcite{DePropris:99}({de~Propris} {et~al.} 1999),
we expect to detect cluster galaxies fainter than $L^*$ at $z=2$ at 3.6
and $4.5\mu$m.  Much of the signal for photometric redshifts, which
increase the contrast of such clusters, derives from the 4000\AA\ break
or from the Balmer decrement. Hence deep complementary optical imaging
is needed, and near-IR imaging is also helpful.  The NOAO Deep
Wide-Field Survey \markcite{Jannuzi:99}(hereafter NDWFS; Jannuzi \& 
Dey 1999) meets this
need with imaging in $B_W, R, I, J,$ and $K_s$, and the 
$9.3\sq^\circ$ NDWFS Bo\"otes
region in particular has both high Galactic and high ecliptic latitude,
providing low backgrounds for {\it Spitzer} imaging. The field center is 
near $\alpha=14^h32^m06^s,
\delta=+34\deg16\arcmin47\arcsec$ (J2000).   Observations of
the Bo\"otes field have also been completed in the radio 
\markcite{deVries:02}({de~Vries} {et~al.} 2002),
in the far-IR with MIPS on {\it Spitzer}, in the near-IR with FLAMINGOS
\markcite{Elston:04}(Elston {et~al.} 2004), and in the X-ray with 
ACIS on {\it Chandra}
\markcite{Murray:04, Kenter:04}(Murray {et~al.} 2004; Kenter {et~al.} 
2004).  Far-UV observations with {\it GALEX} are underway.

The shallow survey team plans to address many other astrophysical objectives using these datasets, ranging from identifications and size estimates for high ecliptic
latitude asteroids from $8\mu$m thermal
emission, to foreground subtraction for detection of fluctuations in the $1 -3\mu$m cosmic background due to Ly$\alpha$ from Population~III objects at $z\sim 15$
\markcite{Cooray:04}(\eg Cooray {et~al.} 2004).

\section{Observing Strategy and Data Processing}

To achieve reasonable reliability, the survey design follows Clyde 
Tombaugh's admonition
to obtain at least three observations at each location.  This results 
in a total of 17,076
separate $5\arcmin \times 5 \arcmin$ images in the four IRAC bands.  
The 30s frame time used provides
background-limited or nearly background-limited data in all four IRAC
channels, and dominates the $\sim10$s needed to reposition by one IRAC
field.

The 45 Astronomical Observation Requests (AOR's) making up the 
shallow survey were
executed during the 3rd
IRAC campaign, on UT 2004 Jan 10 $-$ 14.  The survey area was split
into 15 ``groups'', each of which was observed three times with at least
2 hours between observations to ensure that asteroids
could be reliably identified.  The observation strategy incorporates
several elements to facilitate self-calibration of the data
\markcite{Fixsen:00}(\eg Fixsen, Moseley, \& Arendt 2000): (i) {\em 
dithering}: each AORs starts
at a different point in the small-scale cycling dither table; (ii) 
{\em offsets}:  relative to the first AOR
of each group, subsequent AOR revisits to a group are shifted by 1/3rd
of the 290\arcsec\ step size used by our mapping; and (iii) {\em
cadencing}: for large groups which are rectangular
grids, revisits cover the same area with a
different step size.  For example, the first AOR of
the first group uses an $8 \times 12$ grid map, while the revisits use
a $9 \times 12$ and a $8 \times 13$ grid map.  These strategies
increase the intra-pixel correlations, providing a better self-calibration 
figure of merit (\markcite{Arendt:00}Arendt, Fixsen, \& Moseley 2000).


A version 1 reduction was created using 
the basic methodology of the IRAC pipeline
(IRAC Data Handbook, ver.~1.0; {\tt
http://ssc.spitzer.caltech.edu/irac/dh/}), with a
few additional steps. Residual pattern noise in the images was removed
by subtracting a median combination of the images for each AOR.  Independent
median images were constructed for each amplifier of each
detector, with zero DC level to preserve the ``sky" value of
the images.  A remaining ``jailbar" pattern in the images (which was
worst for $5.8\mu$m) was removed by adding a small constant to the
image from each amplifier so that a detector's four output
amplifiers have equal median ``sky" values.  A correction for some
of the known IRAC array artifacts associated with bright stars
(mux-bleed and column pulldown) was applied.

Additional processing was needed for the $5.8\mu$m data because these
frames suffered obvious heightened systematic effects. A secondary
median-combined delta-dark frame was generated and applied for the
first $5.8\mu$m image taken in each AOR.  Finally an additive offset
was applied to normalize the sky values in each individual $5.8\mu$m
image to the median of all $5.8\mu$m images.  This step removed any
information at $5.8\mu$m about structure larger than one IRAC field,
but was needed to remove systematic trends within and between AORs.

The processed frames were combined into a single image for each channel
using the SSC MOPEX software, after some experimentation with cosmic ray
rejection parameters and comparision with alternate reduction software.
The output pixel scale of the mosaiced image is $\sqrt{2}$ smaller
than the input pixel scale, and the output orientation was rotated
$\sim45\deg$, following the recommendation of 
\markcite{vanDokkum:00}van Dokkum {et~al.} (2000).
Fig.~\ref{figure.plate} (Plate~X) illustrates the full shallow survey
mosaic at $4.5\mu$m, with various interesting sources highlighted.

\section{Photometry}

Source detection and photometry was carried out using SExtractor
\markcite{Bertin:96}(Bertin \& Arnouts 1996) in areas with at
least 3 exposures (listed in Table 1).
Unless otherwise noted,
Vega-relative magnitudes are used throughout.  
Photometry was measured in 3\arcsec\ and  6\arcsec\ diameter
apertures, and using the SExtractor
``MagAuto'' (roughly an isophotal magnitude, indicated
by the subscript ``auto'').  Aperture
magnitudes were corrected to the 10 pixel radius (24\farcs4 diameter)
calibration diameter using values determined by the IRAC instrument
team.   For stars, SExtractor MagAuto values agreed with these 
aperture-corrected values
to $\sim0.1$ mag.

Table 1 gives the survey depths from the standard deviation of
randomly-located 3\arcsec\ and 6\arcsec\ apertures, and  saturation
limits from inspection of bright star radial profiles in single
exposure mosaics.  While we believe the sources detected at $5\sigma$
in 3\arcsec\ apertures to be generally reliable, the more conservative
6\arcsec\ values are used hereafter unless otherwise specified.

\begin{figure}[!t]
\begin{center}
\plotone{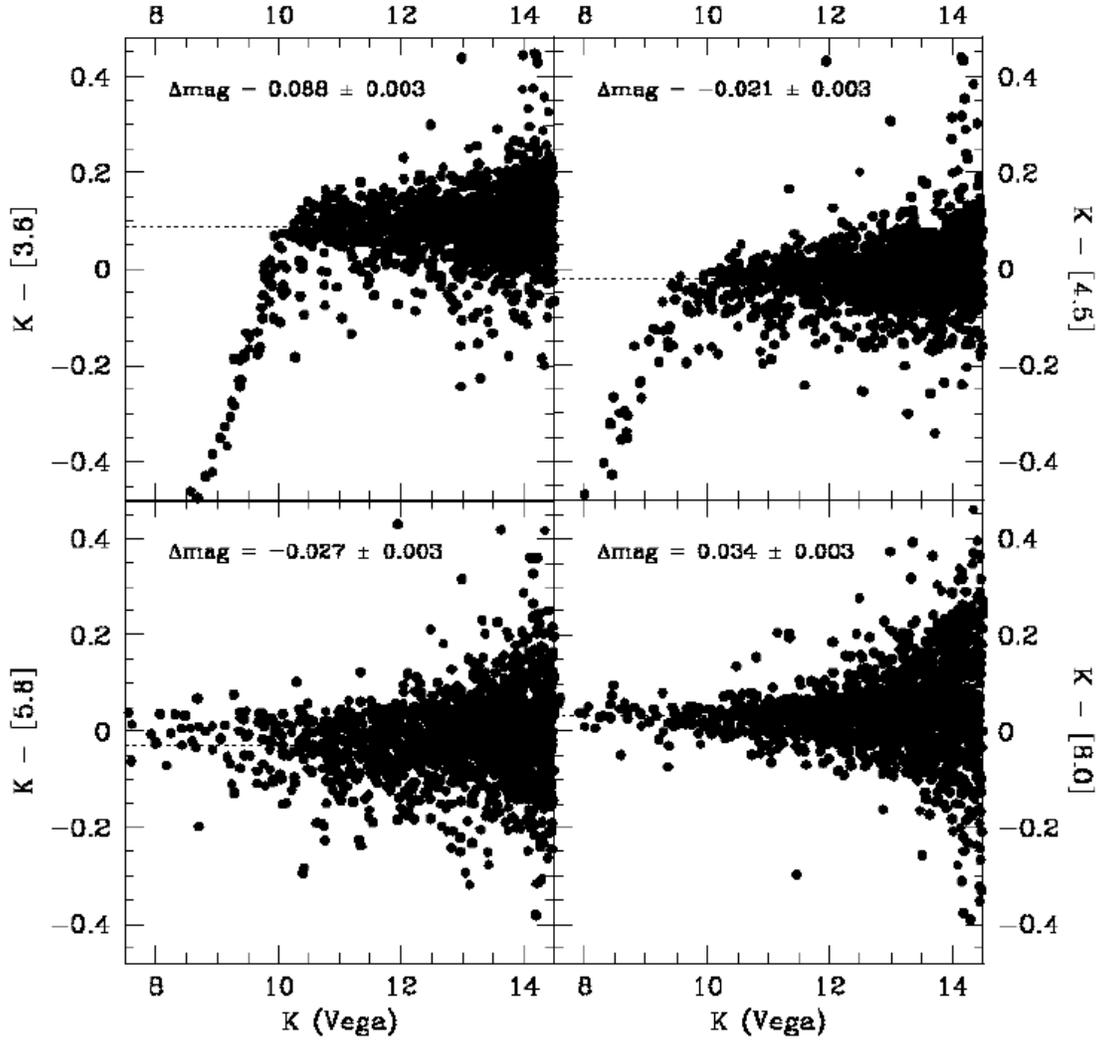}
\end{center}

\caption{$K -$ IRAC versus $K$ for 2MASS counterparts with $J - K
\leq 0.5$.   IRAC photometry is measured in 6\arcsec\ diameter apertures,
corrected to 10 pixel radius (24.4\arcsec\ diameter) apertures.
The horizontal dotted lines
show the average of the distribution for $10 < K < 13$.
Saturation effects are apparent at 3.6 and $4.5\mu$m for sources with 
$K < 10$.}

\label{figure.2mass}
\end{figure}

The IRAC zeropoints, provided by the SSC, have a reported absolute
calibration accuracy better than 10\%. 
As a check, the IRAC sources were matched with 2MASS observations of the field,
yielding $\approx 12,000$ matches per IRAC band.  Sources with $J -
K \leq 0.5$ are expected to be spectral-type K0V (or G2III) and earlier
stars (Johnson 1966; Bessell \& Brett 1988); the bluest nearby galaxies
have $J - K > 0.7$ \markcite{Jarrett:03}(Jarrett {et~al.} 2003).  Most
stars should have close to zero color in IRAC bands (in the Vega
system), as these wavelengths sample the Rayleigh-Jeans tail.
Fig.~\ref{figure.2mass} plots 2MASS to IRAC colors for the 2MASS $J - K
\leq 0.5$ sources.  These sources average $K - [3.6] = 0.09$, while the
other IRAC bands are within 0.03~mag of the expected zero color.
Bessell \& Brett (1988) find that stars of spectral type B8V through K0V
have $K - L$ colors of $-0.03$ to $+0.06$ and $K - M$ colors of $-0.05$
to $+0.03$. We conclude that the IRAC zeropoints are accurate at the
10\%\ level.

\section{Star-Galaxy Separation}

\begin{figure}[!t]
\begin{center}
\plotone{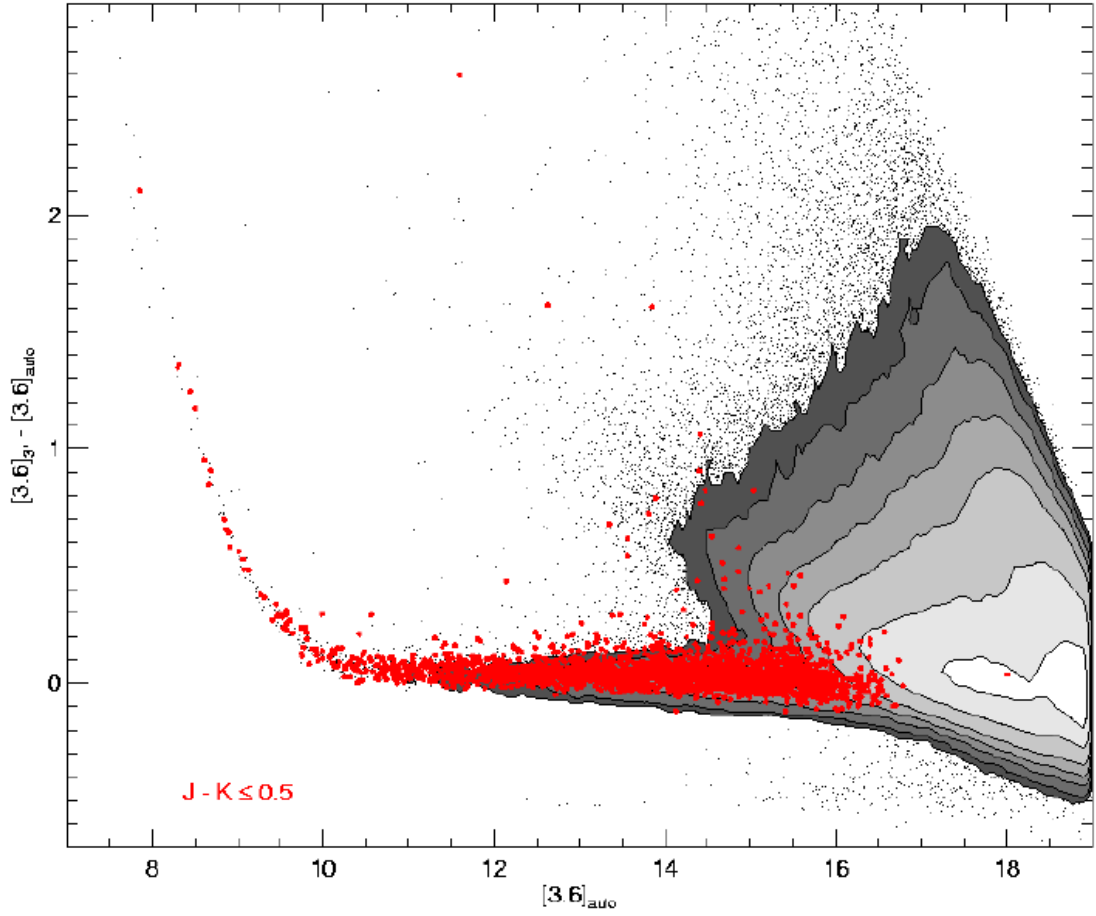}
\end{center}

\caption{Concentration parameter $[3.6]_{3\arcsec} -
[3.6]_{\rm auto}$ versus $[3.6]_{\rm auto}$, where
$[3.6]_{3\arcsec}$ is the SExtractor-derived flux of
sources in 3\arcsec\ diameter apertures, corrected to 24.4\arcsec\
apertures, and $[3.6]_{\rm auto}$ is the
SExtractor-derived MagAuto.   
Progressively lighter shaded contours separate regions at 
surface densities of 
$[2,5,10,20,50,100] \times 10^3$ objects per mag per mag.
Red points indicate 2MASS counterparts with
$J - K \leq 0.5$, which is a robust selection criterion for Galactic stars.}

\label{figure.stargxy}

\end{figure}

Stars dominate the number counts for $m < 14$ in all IRAC bands
(\markcite{Fazio:04b}Fazio {et~al.} 2004), so
accurate identification of stars is necessary to avoid large errors in
bright galaxy counts.  Fig.~\ref{figure.stargxy} shows that morphological
identification of stars with $[3.6] < 15$ is practical.  
Sources with $-0.25 < [3.6]_{3\arcsec} - [3.6]_{\rm auto} <
0.2$ and $[3.6]_{\rm auto} < 15.25$ were counted as stars by 
Fazio {et~al.} (2004), while those
with $[3.6]_{3\arcsec} - [3.6]_{\rm auto} < -0.25$ were rejected as probable
cosmic rays.   For unsaturated 4.5, 5.8, or $8\mu$m sources with $[3.6] < 
10.25$, a similar concentration criterion was used in the relevant 
filter.  Because $3.6\mu$m classification was used,
the effective area for number counts at $m < 15$ is reduced to 8.06,
8.50, and $8.07\sq^\circ$ at 4.5, 5.8, and $8\mu$m.  Objects with
$m < 15$ at 4.5, 5.8, and $8\mu$m but with $[3.6] \ge 15.25$ were counted
as galaxies.  For each wavelength, the brightest two magnitude bins
of objects classified as galaxies by these criteria were individually
inspected, and in some cases reclassified as stars or artifacts. At
$5.8\mu$m, the fraction of bright artifacts was significant.

In the present paper we identify stars by concentration for
$[3.6] < 15.0$, and via optical + IRAC photometry for fainter objects.  
A clear sequence
of stars with 
$(B_W - I) > 2(I - [3.6]) - 1.65$ 
is evident (Fig.~\ref{figure.colcol}a), analogous to the
$BIK$ sequence identified by \markcite{Huang:97}Huang {et~al.} 
(1997).  However, $\sim0.5$ mag redder in $I - [3.6]$ 
than this criterion, a secondary sequence appears, containing roughly $40\%$ as
many stars.  These are likely to be giant stars 
(Johnson 1966, Bessell \& Brett 1988), and we label morphologically compact
$3.6\mu$m sources with $-1.65 > (B_W - I) - 2(I - [3.6]) > -3.35$ using 
yellow star symbols 
elsewhere in Fig.~\ref{figure.colormag} and \ref{figure.colcol}.
Together dwarf and giant stars account for $\approx17\%$ of 3.6
and $4.5\mu$m sources in this survey.

\section{Color-Magnitude and Color-Color Relations}

To assess the general characteristics of the survey data, we have
constructed a variety of color-magnitude and color-color plots,
examples of which are illustrated in Fig.~\ref{figure.colormag} and
\ref{figure.colcol}.  In the released NW 1.2 $\sq^\circ$ of the NDWFS
Bo\"otes field, optical photometry was measured for IRAC shallow survey
sources, and limits of $I = 24$ and $B_W = 26.7$ were used.  
Photometry for objects with $I < 16$ or $B_W < 17$ 
was not used because of saturation effects.  
Absolute astrometry for the shallow survey
(calibrated to 2MASS) and the NDWFS (calibrated to the USNO-A2.0) in
this region agrees to within 0.3\arcsec\ rms.
  
The diagonal edges which bound the data at lower left and upper right
in Fig.~\ref{figure.colormag} arise from the saturation and $2\sigma$
limits respectively in the bluer band.  Fig.~\ref{figure.colormag}b shows
a similar feature bounding the red symbols, corresponding to the
$[3.6] < 15$ morphological classification limit, and this limit appears
as a vertical edge in Fig.~\ref{figure.colormag}d.
The number of objects redder than
the $2\sigma$ limits are listed as a function of magnitude along the top axes
in Fig.~\ref{figure.colormag}.  Each such red object was inspected visually 
to confirm its reality when there were less than 100 in a magnitude 
bin.  For magnitude bins with 100 or more objects redder than the limit, a 
representative subsample was inspected and used to provide a rough correction
($< 10\%$ in all cases) to remove spurious objects from the listed numbers.

\section{Initial Results}

\begin{figure}
\epsscale{0.9}
\plotone{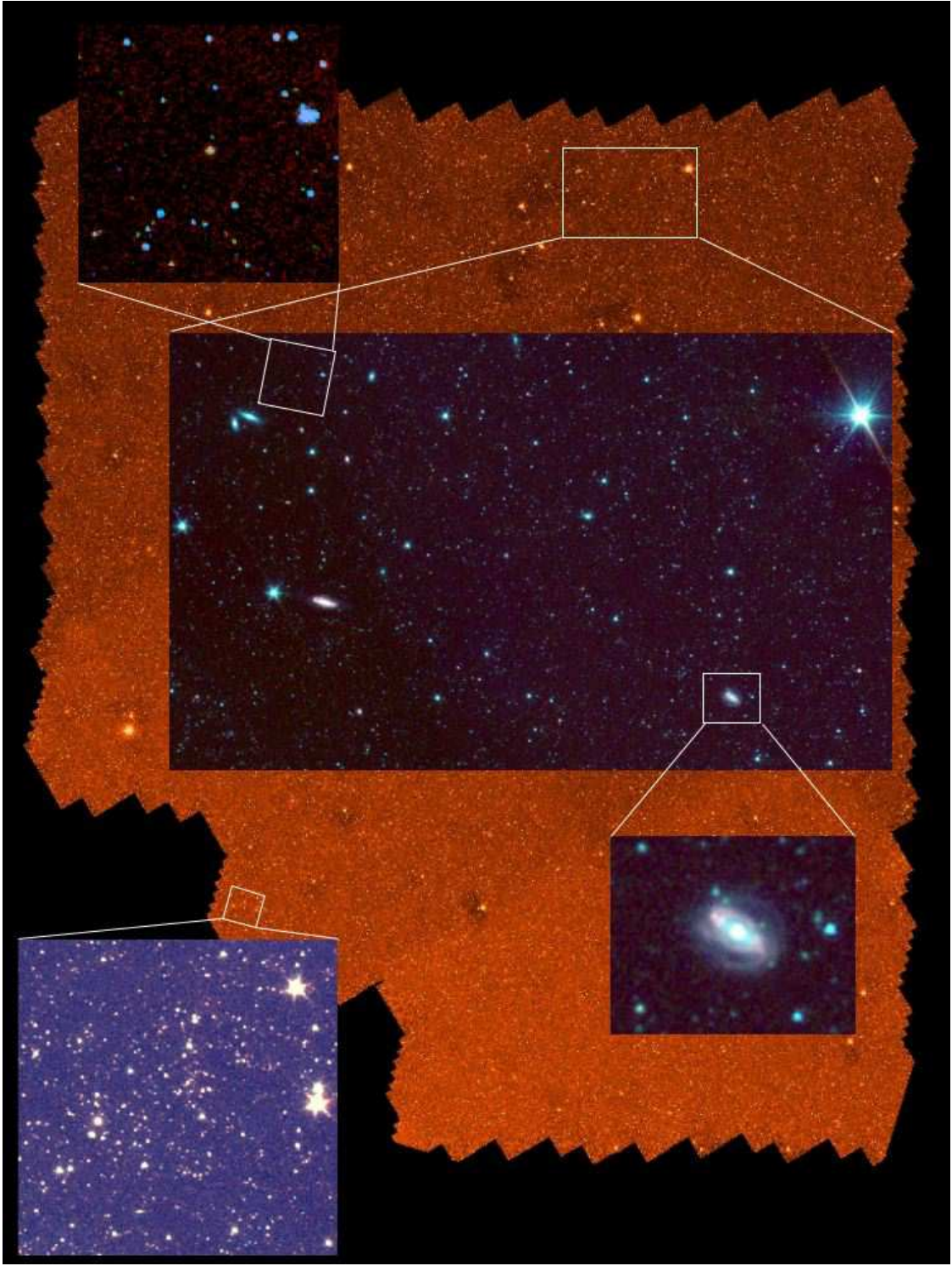}
\caption{Mosaic 4.5 $\mu$m image of the complete IRAC shallow survey,
with N up and E left, subtending $\approx 2.92\deg \times 3.54\deg$.
Insets highlight the galaxy group UGC9315 (at left in large central inset), NGC~5646 (lower right), the extreme $8\mu$m to $I$ flux ratio object
IRAC J142939.1+353557 (upper left), and a region with a high density
of sources with red 3.6$\mu$m - 4.5$\mu$m colors, suggestive of a $z >
1$ cluster (lower left).  Blue, green, and red in the insets correspond to 
3.6, 4.5, and $8\mu$m, except for the lower left where red is $4.5\mu$m
and green is the average of 3.6 and $4.5\mu$m.}
\label{figure.plate}
\end{figure}

An initial attempt at identifying $z > 1$ clusters used
$F_\nu(4.5)/F_\nu(3.6) > 1$, based on the rest frame $1.6\mu$m bump
moving from the 3.6 to the $4.5\mu$m channel.  Fig.~\ref{figure.plate}
illustrates a $400\arcsec \times 400\arcsec$ region where the 
density of such objects
is particularly high.  Spectroscopic confirmation of such candidate
clusters is clearly necessary.

\begin{figure}[!t]
\begin{center}
\epsscale{1}
\plotone{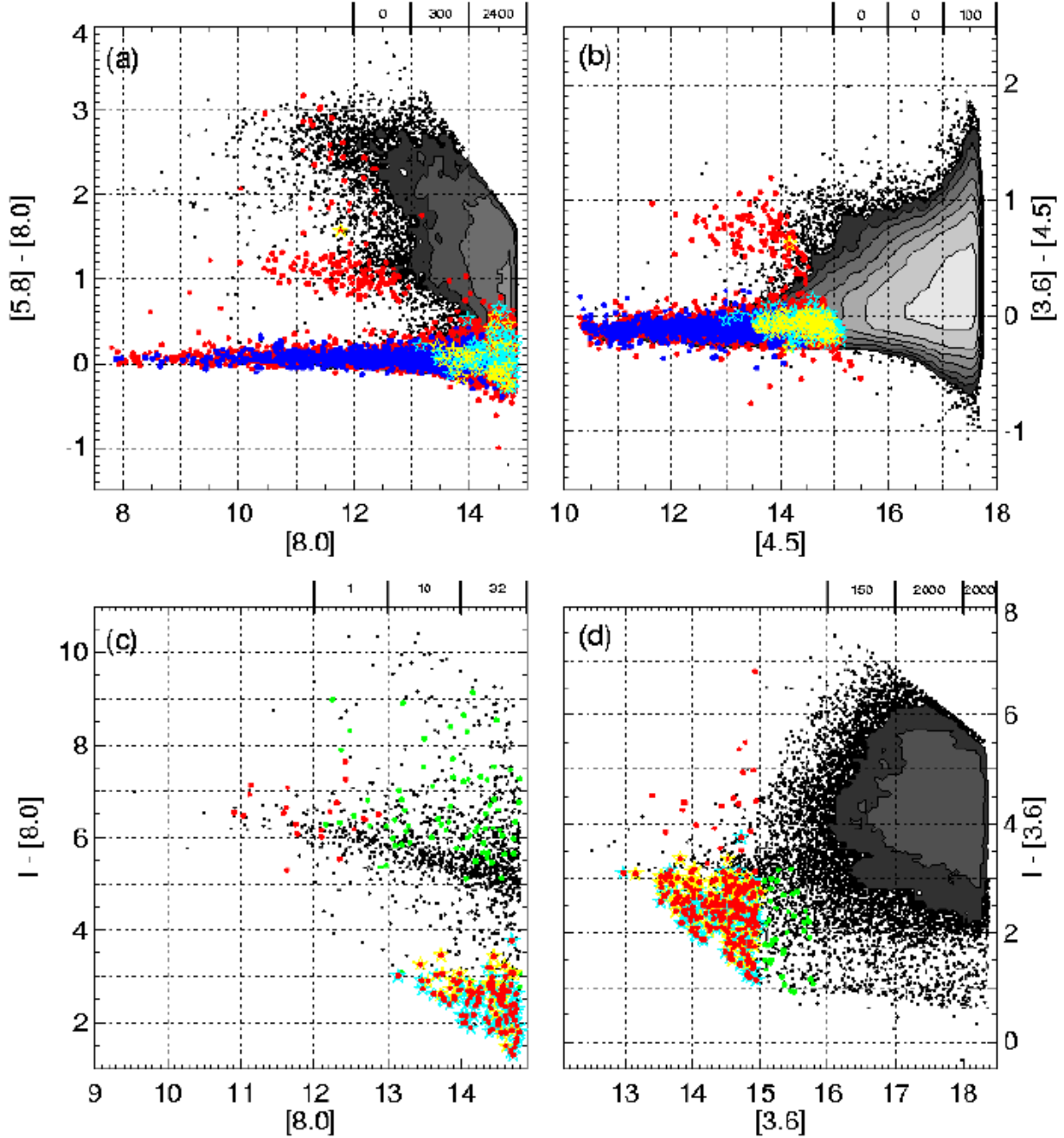}
\end{center}

\caption{Color-magnitude relations for IRAC shallow survey and released
NDWFS Bo\"otes data. Objects were required to be fainter than the
saturation limits and brighter than the 6\arcsec\ $5\sigma$ limits in
Table~1 in the selection band (horizontal axis), and above $2\sigma$ in
the bluer band in order to be plotted.  The numbers of objects redder
than the $2\sigma$ limits are listed along the top axes as a function of
magnitude.  Progressively lighter shaded contours separate regions at 
surface densities of 
$[1,2,5,10,20,50,100] \times 10^3$ objects per mag per mag.
Objects classified as stars based on the $3.6\mu$m
concentration are plotted as red, while objects with 2MASS $J - K \leq 
0.5$ are plotted as blue.  Objects identified as dwarf stars in Fig. 5a
appear as blue star symbols, while giant stars from Fig. 5b are shown
as yellow stars.  Morphological stars from NDWFS $I$ data (SExtractor
stellarity index $> 0.95$) are plotted in green.}

\label{figure.colormag}
\end{figure}

\begin{figure}[!t]
\begin{center}
\epsscale{1}
\plotone{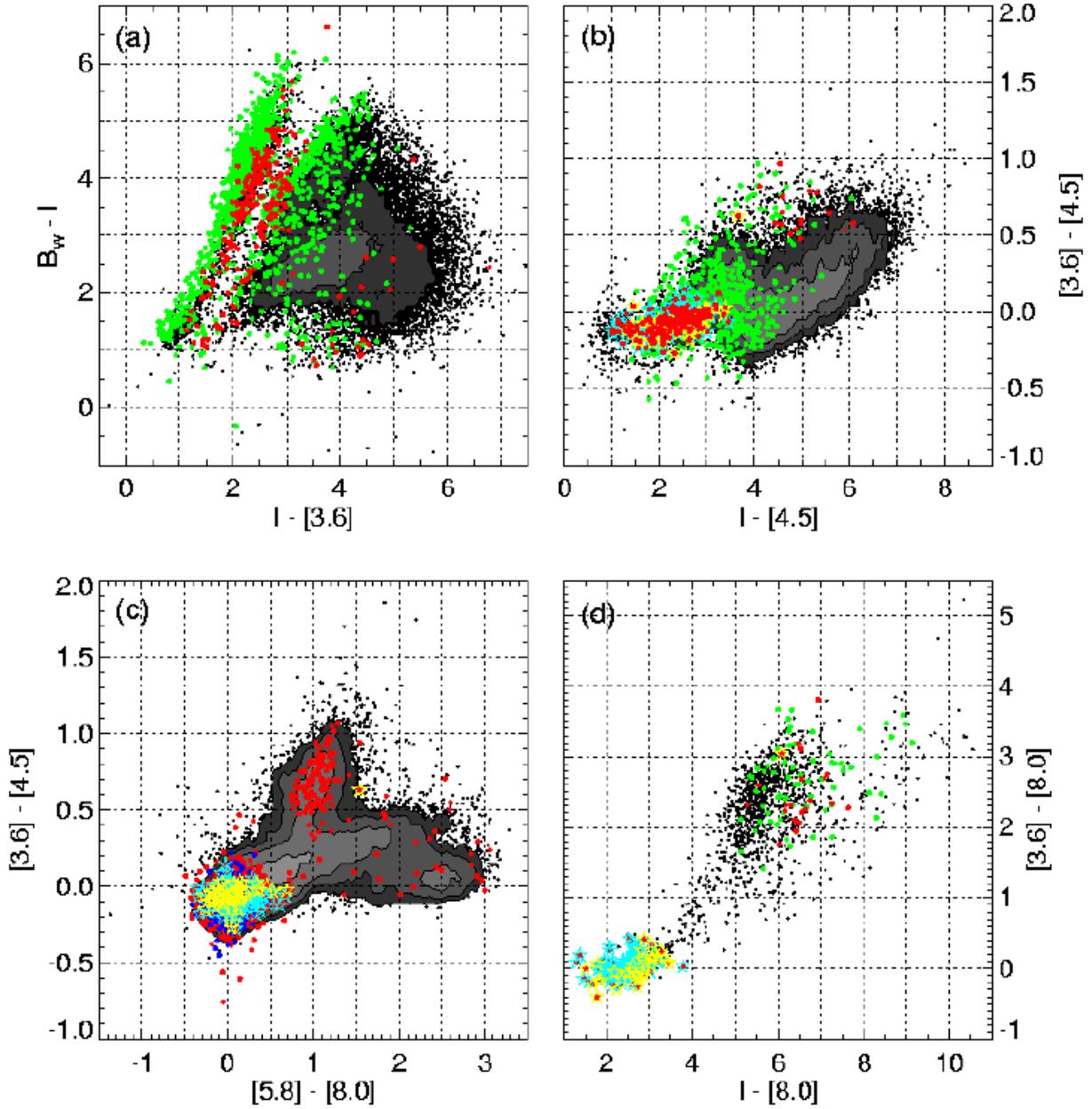}
\end{center}
\caption{Color-color relations for IRAC shallow survey and released
NDWFS Bo\"otes data. Objects are detected at or above $5\sigma$ in all
bands plotted.  Shaded contour levels and 
symbol colors are the same as in figure 4.  }

\label{figure.colcol}
\end{figure}

The nearly horizontal edge at $I - [3.6] \sim 1$ in Fig.~\ref{figure.colormag}d
appears to arise from an absence of stars bluer than this color
in this field (cf Fig.~\ref{figure.colcol}a).  
A stellar sequence is apparent in most panels in Fig.~\ref{figure.colormag} and
\ref{figure.colcol}, but a sequence of
compact $3.6\mu$m sources (red symbols) appears in
Fig.~\ref{figure.colormag}a, \ref{figure.colormag}b, and
\ref{figure.colcol}c with different colors than the dwarf and giant
stars noted in \S 5.
Preliminary spectroscopy of unresolved sources outside the stellar color loci
suggests they are a mixture of broad-lined
quasars and starburst galaxies.  We find $\approx 300$
such IRAC-selected objects in the released 1.2 $\sq^\circ$ 
corner of the NDWFS.  It is premature
to assess what fraction of these candidates are quasars, but it is noteworthy
that the Sloan Digital Sky Survey finds a surface density of
only 15 quasars$/\sq^\circ$ with $i^* \leq 19.1$
(Richards \etal\ 2002).
                          
\begin{figure}[!t]
\plotfiddle{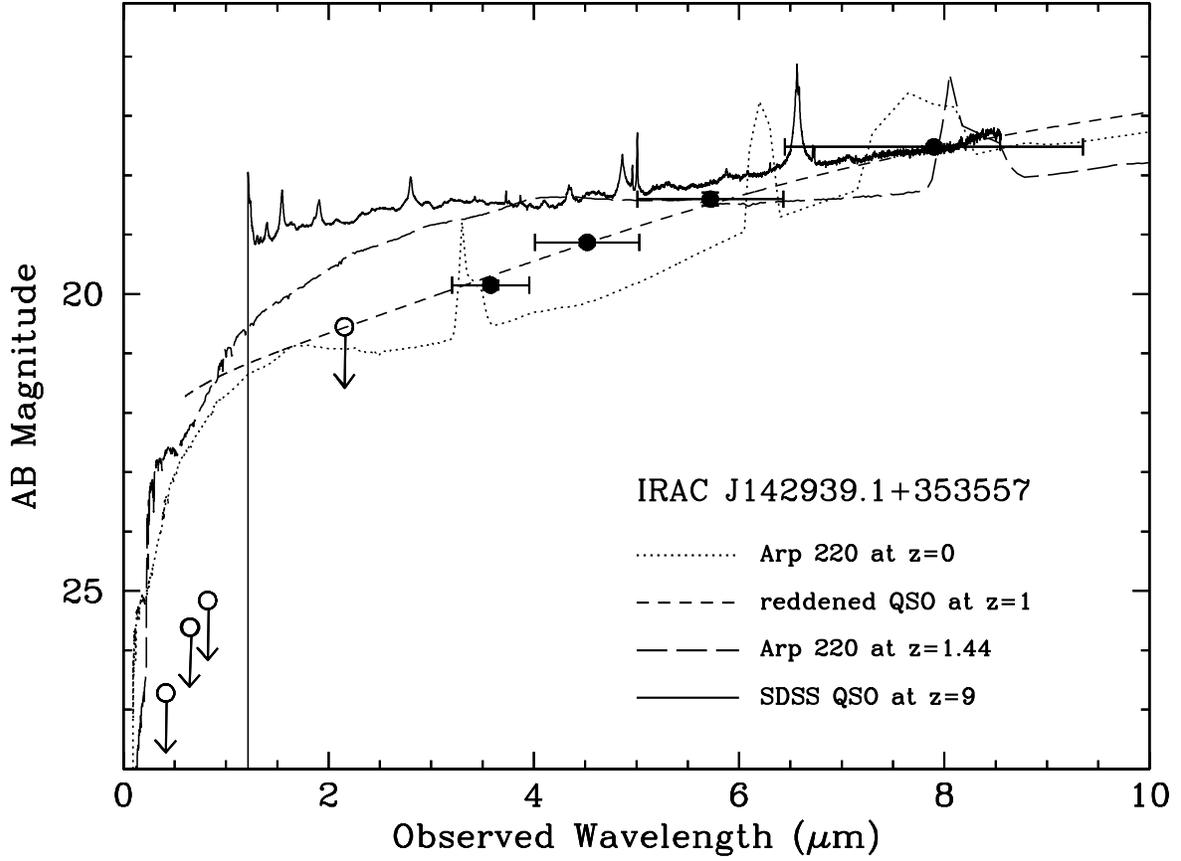}{3.5in}{-90}{60}{60}{-230}{350}

\caption{Observed fluxes for IRAC J142939.1+353557, an object with an extreme ratio of $8\mu$m to $I$ flux.  
Models are explained in the text.}

\label{figure.redguy}
\end{figure}

Finally, we give an illustration of the possibilities and challenges of
identifying rare objects in the IRAC shallow survey.  Using the
released $1.2\sq^\circ$ portion of the NDWFS,  a search was carried out
for objects bright enough at $8\mu$m for IRS short-low spectroscopy
(taken as 0.35 mJy), but which are extremely challenging for optical
spectroscopy (taken as $I \geq 24$).  The vast majority of the 678
sources in this region with $8\mu$m fluxes above 0.35 mJy are real, but
selecting for unusual colors enhances the probability of spurious
sources.  Visual examination of the 19 candidates with $I \geq 24$
showed most were attributable to cosmic rays or scattered light from
bright stars in the $8\mu$m data.  We therefore also required that they also be
identifiable at 3.6 or 4.5 $\mu$m.  This reduced the number of
candidates to 4, of which the faint or absent $I$ flux for 2 could be
attributed to problems with the $I$ data.  The remaining two objects
are deemed reliable.  Fig.~\ref{figure.redguy} shows the SED for one of
these (the other is similar), together with some model spectra
normalized at $8\mu$m.

How can we account for the properties of these objects?  
Their $8\mu$m to $0.8\mu$m
flux ratio is $> 500$, a spectral index of $> 2.7$.  One possibility
is that these are relatively nearby starburst galaxies, whose
$8\mu$m flux is enhanced by $7.7\mu$m PAH emission.  The $z=0$ 
model spectrum of Arp~220 from \markcite{Devriendt:99}Devriendt, 
Guiderdoni, \&  Sadat (1999) illustrates this, but
fails to match the observed $I$ flux limit,
needing an additional $A_I \sim 2$, equivalent to 
$A_V \sim 4$.  The same 
model at $z=1.4$ places rest
$3.3\mu$m emission at $8\mu$m observed, with a similar
mismatch at $I$.  However in this case the $A_V$ equivalent is only $\sim
1$ mag, because observed $I$ samples rest $0.32\mu$m.  The luminosity
would be comparable to that of the ULIRG HR10 (ERO~J164502+4626.4),
i.e. in excess of $10^{12} L_\odot$ \markcite{Dey:99}(Dey {et~al.} 1999).
The dashed line in
Fig.~\ref{figure.redguy} shows a quasar at $z=1$ with $A_V\sim 3$ 
from \markcite{Polletta:00}Polletta {et~al.} (2000) which 
matches the the IRAC data well,
but again additional reddening is needed to match the $I$ limit.  
This spectrum matches the reddest AGN found by 2MASS.  A final, less
likely possibility is a quasar at $z > 6$ in which the Ly$\alpha$ forest 
suppresses observed light
below $1\mu$m, as illustrated by the SDSS composite quasar spectrum
in Fig.~\ref{figure.redguy} from 
\markcite{VandenBerk:01}{Vanden~Berk} {et~al.} (2001).  However
\markcite{Fan:03}Fan {et~al.} (2003) find only $\sim1$ such quasar 
per 500 square degrees.

Followup observations with {\it Spitzer's} Infrared Spectrograph 
are likely to reveal which of these
possibilities is correct, or whether these objects represent some new
phenomenon.  The large volume probed by the IRAC shallow survey data 
will allow the discovery of many more objects with unusual colors.

\acknowledgments

We thank H. Spinrad and S. Dawson for providing 
spectroscopy of IRAC-selected quasar candidates.
This work is based on observations made with the {\it Spitzer Space
Telescope}, which is operated by the Jet Propulsion Laboratory,
California Institute of Technology  under NASA contract 1407. Support
was provided by NASA through an award issued by
JPL/Caltech.  The NDWFS would not have been 
possible without 
support from NOAO, which is 
operated by the Association of Universities for Research in Astronomy 
under a cooperative agreement with the National Science 
Foundation (NSF).  This publication makes use of data products from the Two
Micron All Sky Survey, which is a joint project of the U. of
Massachusetts and the Infrared Processing and Analysis
Center/Caltech, funded by NASA and the NSF.


\clearpage
\begin{deluxetable}{lcccccc}
\tablecaption{Properties of the IRAC Shallow Survey}
\tablehead{
\colhead{$\lambda$} &
\colhead{Area} &
\colhead{5$\sigma$ 3"} &
\colhead{5$\sigma$ 3"} &
\colhead{5$\sigma$ 6"} &
\colhead{5$\sigma$ 6"} &
\colhead{Saturation} \\
\colhead{($\mu$m)} &
\colhead{($\sq^\circ$)} &
\colhead{($\mu$Jy)} &
\colhead{(Vega mag)} &
\colhead{($\mu$Jy)} &
\colhead{(Vega mag)} &
\colhead{(Vega mag)}}
\startdata
3.6 & 8.55 &  6.4 & 19.1 & 12.3 & 18.4 & 10.0 \nl
4.5 & 8.53 &  8.8 & 18.3 & 15.4 & 17.7 &  9.8 \nl
5.8 & 8.54 &  51  & 15.9 & 76   & 15.5 &  7.5 \nl
8.0 & 8.54 &  50  & 15.2 & 76   & 14.8 &  7.7 \nl
\enddata
\end{deluxetable}

\clearpage

\begin{thebibliography}{}

\bibitem[Arendt, Fixsen, \& Moseley 2000]{Arendt:00}
Arendt, R.~G., Fixsen, D.~J., \& Moseley, S.~H. 2000, \apj, 536, 500

\bibitem[Bertin \& Arnouts 1996]{Bertin:96}
Bertin, E. \& Arnouts, S. 1996, \aaps, 117, 393

\bibitem[Bessell \& Brett 1988]{Bessell:88}
Bessell, M.S. \& Brett, J.M. 1988, PASP, 100, 1134


\bibitem[Cooray {et~al.} 2004]{Cooray:04}
Cooray, A., Bock, J. J., Keatin, B., Lange, A. E., \& Matsumoto, T. 
2004, ApJ, 606, 611

\bibitem[{de~Propris}, Stanford, Eisenhardt,  Dickinson, \& Elston 
1999]{DePropris:99}
{de~Propris}, R., Stanford, S.~A., Eisenhardt, P.~R.~M., Dickinson, 
M., \&  Elston, R. 1999, \aj, 118, 719

\bibitem[{de~Vries}, Morganti, R\"ottgering,  Vermeulen, 
{van~Breugel}, Rengelink, \& Jarvis 2002]{deVries:02}
{de~Vries}, W.~H., Morganti, R., R\"ottgering, H.~J.~A., Vermeulen, 
R.,  {van~Breugel}, W., Rengelink, R., \& Jarvis, M.~J. 2002, \aj, 
123, 1784

\bibitem[Devriendt, Guiderdoni, \&  Sadat 1999]{Devriendt:99}
Devriendt, J.~E.~G., Guiderdoni, B., \& Sadat, R. 1999, \aap, 350, 381

\bibitem[Dey, Graham, Ivison, Smail, Wright, \&  Liu 1999]{Dey:99}
Dey, A., Graham, J.~R., Ivison, R.~J., Smail, I., Wright, G.~S., \& 
Liu, M.~C.  1999, \apj, 519, 610

\bibitem[Eisenhardt \& Wright 2003]{Eisenhardt:03}
Eisenhardt, P.~R. \& Wright, E.~L. 2003, SPIE, 4850, 1050

\bibitem[Elston {et~al.} 2004]{Elston:04}
Elston, R. {et~al.} 2004, in preparation

\bibitem[Fan {et~al.} 2003]{Fan:03}
Fan, X. {et~al.} 2003, \aj, 125, 1649

\bibitem[Fazio {et~al.} 2004]{Fazio:04b}
Fazio, G.~G. {et~al.} 2004, \apjs, this volume

\bibitem[Fixsen, Moseley, \& Arendt 2000]{Fixsen:00}
Fixsen, D.~J., Moseley, S.~H., \& Arendt, R.~G. 2000, \apj, 128, 651

\bibitem[Hogg, Neugebauer, Cohen, Dickinson, Djorgovski,  Matthews, 
\& Soifer 2000]{Hogg:00}
Hogg, D.~W., Neugebauer, G., Cohen, J.~G., Dickinson, M., Djorgovski, 
S.~G.,  Matthews, K., \& Soifer, B.~T. 2000, \aj, 119, 1519

\bibitem[Huang, Cowie, Gardner, Hu, Songaila, \&  Wainscoat 1997]{Huang:97}
Huang, J., Cowie, L.~L., Gardner, J.~P., Hu, E.~M., Songaila, A., \& 
Wainscoat,  R.~J. 1997, \apj, 476, 12

\bibitem[Jannuzi \& Dey 1999]{Jannuzi:99}
Jannuzi, B.~T. \& Dey, A. 1999, in {\it Photometric Redshifts and 
High-Redshift  Galaxies}, ed. R.~Weymann, L.~Storrie-Lombardi, 
M.~Sawicki, \& R.~Brunner,  Vol. 191 (San Francisco: ASP Conference 
Series), 111

\bibitem[Jarrett, Chester, Cutri, Schneider, \&  Huchra 2003]{Jarrett:03}
Jarrett, T.~H., Chester, T., Cutri, R., Schneider, S.~E., \& Huchra, 
J.~P.  2003, \aj, 125, 525

\bibitem[Johnson 1966]{Johnson:66}
Johnson, H.~L. 1966, \araa, 4, 193

\bibitem[Kenter {et~al.} 2004]{Kenter:04}
Kenter, A. {et~al.} 2004, in preparation


\bibitem[Murray {et~al.} 2004]{Murray:04}
Murray, S.~S. {et~al.} 2004, in preparation

\bibitem[Polletta, Courvoisier, Hooper, \&  Wilkes 2000]{Polletta:00}
Polletta, M., Courvoisier, T.~J., Hooper, E.~J., \& Wilkes, B. 2000, 
\aap, 362,  75

\bibitem[{et~al.} 2002]{Richards:02}
Richards, G.T. \etal 2002, \aj, 123, 2945


\bibitem[Sato {et~al.} 2003]{Sato:03}
Sato, Y. {et~al.} 2003, \aap, 405, 833

\bibitem[van Dokkum, Franx, Fabricant, Illingworth,  \& Kelson 
2000]{vanDokkum:00}
van Dokkum, P.~G., Franx, M., Fabricant, D., Illingworth, G.~D., \& 
Kelson,  D.~D. 2000, \apj, 541, 95

\bibitem[{Vanden~Berk} {et~al.} 2001]{VandenBerk:01}
{Vanden~Berk}, D.~E. {et~al.} 2001, \aj, 122, 549


\end{thebibliography}
\end{document}